# Tsunami Wave Runup on Coasts of Narrow Bays

**Narcisse Zahibo[1], Efim Pelinovsky[1,2], Vladimir Golinko[3], Nataly Osipenko[3]**

[1] Département de Physique, Université des Antilles et de la Guyane, UFR Sciences, Campus de Fouillole, 97159 Pointe à Pitre Cedex, Guadeloupe, France; Email: narcisse.zahibo@univ-ag.fr

[2] Laboratory of Hydrophysics and Nonlinear Acoustics, Institute of Applied Physics, 46 Uljanov Str., Nizhny Novgorod, 603950, Russia; Email: pelinovsky@hydro.appl.sci-nnov.ru

[3] Department of Applied Mathematics, Nizhny Novgorod Technical University, 24 Minin Str., Nizhny Novgorod, 603950, Russia

January 31, 2005

Version 1

**Abstract**

The runup of tsunami waves on the coasts of the barrow bays, channels and straits is studied in the framework of the nonlinear shallow water theory. Using the narrowness of the water channel, the one-dimensional equations are applied; they include the variable cross-section of channel. It is shown that the analytical solutions can be obtained with use of the hodograph (Legendre) transformation similar to the wave runup on the plane beach. As a result, the linear wave equation is derived and all physical variables (water displacement, fluid velocity, coordinate and time) can be determined. The dynamics of the moving shoreline (boundary of the flooding zone) is investigated in details. It is shown that all analytical formulas for the moving shoreline can be obtained explicitly. Two examples of the incident wave shapes are analysed: sine wave and parabolic pulse. The last example demonstrates that even for approaching of the crest only, the flooding can appear very quickly; then water will recede relatively slowly, and then again quickly return to the initial state.

*Correspondence:*

Professor Efim Pelinovsky

Laboratory of Hydrophysics and Nonlinear Acoustics,
Institute of Applied Physics,
46 Uljanov Street, Nizhny Novgorod, 603950 Russia
Email: **Pelinovsky@hydro.appl.sci-nnov.ru**
Phone: 007-8312-164839
Fax: 007-8312-365976

# 1. Introduction

To estimate the flooding area of the coastal zone caused by the tsunami waves is essential for tsunami hazard mitigation. Much progress in understanding of the runup processes has been made in applying the one-dimensional nonlinear shallow water theory to the wave runup on the plane beach. It was based on the analytical approach by Carrier and Greenspan (1958), applied the hodograph (Legendre) transformation to transform the initial nonlinear equations into linear wave equations. This method has been used by many authors to analyse the runup of the smoothly nonlinear long waves (Shuto, 1973; Spielfogel, 1976; Pedersen and Gjevik, 1983, Synolakis, 1987, 1991; Pelinovsky and Mazova, 1992; Tadepalli and Synolakis, 1994; Pelinovsky, 1995; Massel and Pelinovsky, 2001; Carrier et al, 2003; Kânoğlu, 2004; Tinti and Tonini, 2005). The applicability of the smooth solutions to the tsunami waves has been checked by Mazova et al (1983) using the existing tsunami database. Approximately 75% of observed tsunami waves in Pacific were not accompanied by wave breaking and, therefore, the shallow water theory is an adequate model to study the tsunami wave runup. The progress in the area of long wave runup is discussed in several special workshops (Liu et al, 1991; Yeh et al, 1996; Yalciner et al, 2003).

The exact estimation of the size of the area flooded by tsunami waves requires to solve two-dimensional shallow-water equations, taking into account the complex geometry of the coastal line contained bays, straits, estuaries, etc. Even for a basin of simple topography, the two-dimensional nonlinear equations cannot be solved analytically using the hodograph transformation as it can be done for the one-dimensional hyperbolic equations. This mathematical difficulty of the solution of hyperbolic equations in two and more dimensions is well known. Several approximations are used to get solutions. Brocchini and Peregrine (1996) studied the nonlinear wave reflection on the plane beach for the case of very small incident angles. The initial problem was split into two different problems: the first described nonlinear one-dimensional offshore dynamics, and the other described approximately linear alongshore dynamics. Golinko and Pelinovsky (1998), Pelinovsky (1993) and Pelinovsky et al (1999) applied another approximation of a very narrow channel. In this case the two-dimensional shallow water equations can be reduced to the equivalent one-dimensional equations. Some particular solutions for specific beach geometry are obtained in cited papers. Present study deals with exact nonlinear solutions of the equivalent one-dimensional equations for more general geometry of coastal zone.

Basic one-dimensional equations of the nonlinear shallow water theory for narrow bays are briefly reproduced in section 2. They are solved with use of the hodograph (Legendre) transformation. The dynamics of the moving shoreline is studied in section 3. It is shown that the exact solution can be found explicitly through the solution of linear problem of the wave runup. The case of runup of the sine wave is analysed in details, and the condition of the wave breaking is obtained (section 4). The runup of the solitary wave is investigated in section 5. It is demonstrated that tsunami wave can approach quickly even the initial disturbance has the one time scale only. Obtained results are summarized in Conclusion.

## 2. Basic model

Tsunami waves generated by strong earthquakes have usually large values of the wavelength to compare with water depth (hundreds km), and their fronts are close to the straight line (quasi-plane waves). The characteristics width of many bays and straits is significantly less (tenth km), and the wave entries in such channels as uniform flow in cross-section. For simplicity we assume the analytical expression for the bottom shape as

$$z(x,y) = -h(x) + f(y). \qquad (1)$$

This geometry is displayed in Fig. 1. If the wave propagates along the *x*-axis, the two-dimensional equations of the nonlinear shallow water theory can be integrated on the cross-section, and the corresponding equations are one-dimensional:

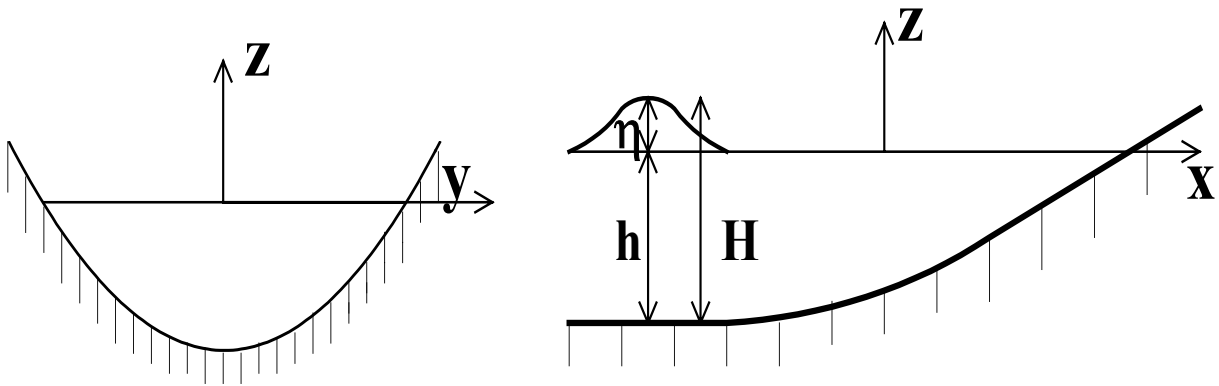

**Fig. 1.** The characterised cross-section and longitudinal projection of the bay

$$\frac{\partial u}{\partial t} + u\frac{\partial u}{\partial x} + g\frac{\partial H}{\partial x} = g\frac{dh}{dx},\tag{2a}$$

$$\frac{\partial S}{\partial t} + \frac{\partial}{\partial x}(Su) = 0,\tag{2b}$$

where $H(x,t) = h(x) + \eta(x,t)$ is the total depth along the channel, $\eta(x,t)$ is the displacement of the water surface, $S(x,t)$ is the area of the cross-section of the channel, and $u(x,t)$ is the mean flow velocity. Integration of (1) makes the system (2) closed, and the solution depends on the beach geometry. In particular, if

$$f(y) = q|y|^m,\tag{3}$$

the function, $S$ is

$$S = \frac{2m}{(m+1)q^{1/m}} H^{1+1/m},\tag{4}$$

where $q$ and $m$ are arbitrary constants. The next approximation is a constant bottom slope of the channel axis; thus

$$h(x) = -\alpha x.\tag{5}$$

The system (2) at these conditions reduces to

$$\frac{\partial u}{\partial t} + u\frac{\partial u}{\partial x} + g\frac{\partial H}{\partial x} = g\frac{dh}{dx}, \qquad \frac{\partial H}{\partial t} + u\frac{\partial H}{\partial x} + \frac{m}{m+1}H\frac{\partial u}{\partial x} = 0,\tag{6}$$

and differs from the "classical" one-dimensional equations for the wave runup on plane beach by constant coefficient $m/(m+1)$, and transforms to them at $m \to \infty$ ($S \sim H$). As a result, the hodograph transformation can be applied for such geometry.

Introducing the Riemann invariants

$$I_\pm = u \pm 2\sqrt{\frac{m+1}{m}gH} + \alpha gt, \tag{7}$$

the system (6) is re-written in the form

$$\frac{\partial I_\pm}{\partial t} + c_\pm \frac{\partial I_\pm}{\partial x} = 0, \tag{8}$$

where characteristic speeds are

$$c_\pm = \frac{3m+2}{4(m+1)}I_+ + \frac{m+2}{4(m+1)}I_- - \alpha gt. \tag{9}$$

Multiplying (8) on the Jacobian $\partial(t,x)/\partial(I_+,I_-)$, assuming that it is not zero (this important question will be discussed in section 4), it can be transformed to

$$\frac{\partial x}{\partial I_\mp} - c_\pm \frac{\partial t}{\partial I_\mp} = 0. \tag{10}$$

The system (10) is nonlinear one due to the dependence $c_\pm$ from $I_\pm$, but it can be reduced to linear, eliminating $x$

$$\frac{\partial^2 t}{\partial I_+ \partial I_-} + \frac{3m+2}{2m(I_+ - I_-)}\left(\frac{\partial t}{\partial I_-} - \frac{\partial t}{\partial I_+}\right) = 0. \tag{11}$$

Let us introduce new arguments:

$$\lambda = \frac{I_+ + I_-}{2} = u + \alpha gt, \tag{12}$$

$$\sigma = \frac{I_+ - I_-}{2} = 2\sqrt{\frac{m+1}{m}gH}. \tag{13}$$

Then, equation (11) takes the form

$$\frac{\partial^2 t}{\partial \lambda^2} - \frac{\partial^2 t}{\partial \sigma^2} - \frac{3m+2}{m}\frac{\partial t}{\partial \sigma} = 0. \tag{14}$$

Extracting time from (12) and substitute

$$u = \frac{1}{\sigma}\frac{\partial \Phi}{\partial \sigma}, \tag{15}$$

equation (14) is re-written in final form

$$\frac{\partial^2 \Phi}{\partial \lambda^2} - \frac{\partial^2 \Phi}{\partial \sigma^2} - \frac{m+2}{m\sigma}\frac{\partial \Phi}{\partial \sigma} = 0. \tag{16}$$

It is convenient to give formulas to determine all physical variables

$$\eta = \frac{1}{2g}\left[\frac{m}{m+1}\frac{\partial \Phi}{\partial \lambda} - u^2\right], \qquad u = \frac{1}{\sigma}\frac{\partial \Phi}{\partial \sigma}, \tag{17}$$

$$x = \frac{\eta}{\alpha} - \frac{m\sigma^2}{4g\alpha(m+1)}, \qquad t = \frac{\lambda - u}{g\alpha}. \tag{18}$$

So, the initial set of nonlinear shallow water equations has reduced to the linear wave equation (16) and all physical variables can be found via $\Phi$ using simple operations. The main advantage of this form is that the moving (unknown) shoreline corresponds now to $\sigma = 0$ (since the total depth H = 0) and, therefore, equation (16) is solved in the half-space $-\infty < \sigma < 0$ with fixed boundary. Such transformation generalizes the Carrier – Greenspan transformation, and reduces to it for plane beach ($m \to \infty$), as it is given in their pioneer work.

## 3. Dynamics of the moving shoreline

The formulas (17) – (18) are implicit and this is a main difficulty to find the wave field analytically. Meanwhile, an important problem for practice – the dynamics of the moving shoreline (the boundary of flooded zone) can be found explicitly for waves generated far from the coast. On large distances from the shore, the wave field is linear, and formulas (17) – (18) give the explicit relations between all variables.

$$\eta_l = \frac{m}{2g(m+1)} \frac{\partial \Phi_l}{\partial \lambda_l}, \qquad u_l = \frac{1}{\sigma_l} \frac{\partial \Phi_l}{\partial \sigma_l}, \qquad (19)$$

$$x_l = -\frac{m\sigma_l^2}{4g\alpha(m+1)}, \qquad t_l = \frac{\lambda_l}{g\alpha}, \qquad (20)$$

where we specially mark all variables as linear variables. Having initial conditions for the zone of the tsunami generation, or knowing the characteristics of the approached tsunami wave, it is easy to find the function $\Phi_l(t_l, x_l)$ or $\Phi_l(\lambda_l, \sigma_l)$. Because functions $\Phi(\lambda, \sigma)$ and $\Phi_l(\lambda_l, \sigma_l)$ are described by the same equation (16), the solution of the wave equation is determined fully, as in the linear approximation, as well as in the nonlinear exact formulation. Considering now $\sigma_l = 0$, we may obtain the wave record at the unmoving shoreline ($x_l = 0$) in the linear approximation (water elevation and fluid velocity), $R_l(t_l) = \eta_l(t_l, 0)$ and $U_l(t_l) = u_l(t_l, 0)$. In the nonlinear formulation, point $\sigma = 0$ corresponds to the moving shoreline, and $R(t) = \eta(\lambda, 0)$ and $U(t) = u(\lambda, 0)$ describe the "real" runup characteristics. Formally, both functions, $R_l(\lambda_l)$ and $R(\lambda)$ are the same, but the argument of the nonlinear runup depends from the velocity. At moment $U = 0$ (maximum runup or rundown), arguments coincide, and, therefore, maximum runup characteristics can be found in the linear formulation. It is an important conclusion for tsunami practice, which is used sometimes with no strong mathematical proof. Moreover, the solution of the linear problem can be used to describe the dynamics of moving shoreline in simple form. According to the last formula (18) for time, the following relation is obtained

$$U(t) = U_l\left(t + \frac{U}{g\alpha}\right), \qquad (21)$$

and "nonlinear" velocity of the moving shoreline is obtained from the "linear" velocity with the variable deformation of time. Such deformation is well-known for the Riemann (simple) wave in gas dynamics deformed in space and time, but here it is valid for moving boundary only. Using (21), the runup displacement can be calculated

$$R(t) = \alpha \int U(t)dt = R_l\left(t + \frac{U}{g\alpha}\right) - \frac{1}{2g}U^2\left(t + \frac{U}{g\alpha}\right). \qquad (22)$$

It is important to mention that the form of the expressions (21) and (22) is not determined by the bay shape, but the "linear" functions, $U_l$ and $R_l$, depend strongly from the bay shape according to the solution of equation (16).

## 4. Runup of sine wave

All theoretical conclusions can be illustrated on the example of the sine wave runup. The particular bounded solution of the wave equation (16) has the following form

$$\Phi(\lambda, \sigma) = Q \frac{J_{1/m}(p\sigma)}{\sigma^{1/m}} \sin(p\lambda), \qquad (23)$$

where $J_{1/m}(\xi)$ is the Bessel function, and $Q$ and $p$ are constants which should be determined. In the linear approximation, the water displacement is

$$\eta_l = \frac{mQp}{2g(m+1)} \frac{J_{1/m}(p\sigma_l)}{\sigma_l^{1/m}} \cos(p\lambda_l). \qquad (24)$$

Using standard asymptotic formulas for the Bessel functions

$$J_\nu(y) \approx \frac{1}{\Gamma(\nu+1)}\left(\frac{y}{2}\right)^\nu \text{ at } y \to 0, \text{ and } J_\nu(y) \approx \sqrt{\frac{2}{\pi y}} \cos\left(y - \frac{\pi\nu}{2} - \frac{\pi}{4}\right) \text{ at } y \to \infty \qquad (25)$$

($\Gamma(\zeta)$ is the Gamma function), we may obtain the shoreline oscillations in the linear theory ($x_l = 0$)

$$\eta_l(t_l) = R_{max} \cos(\omega t_l), \qquad (26)$$

and wave field far from the shoreline

$$\eta(t,x) = A_0 \left(\frac{h}{h_0}\right)^{-\frac{2+m}{4m}} \left\{\sin\left(\omega\left[t - \int dx/c(x)\right]\right) + \sin\left(\omega\left[t + \int dx/c(x)\right]\right)\right\}. \qquad (27)$$

Here $A_0$ is the amplitude of the incident wave with frequency $\omega$ on depth $h_0$,

$$c(x) = \sqrt{\frac{m}{m+1} g h(x)} \qquad (28)$$

is the speed of the linear wave propagation. The asymptotic expression (27) for the far-field is valid for the nonlinear case also, thus we do not use here "linear" symbols. As it is expected, the far-field presents the linear superposition of two waves propagating in the opposite directions, and wave amplitude satisfies to the Green formula due to variability of the depth and bay width.

The parameters, $Q$ and $p$ in (24) can be determined through $A_0$ and $\omega$, and then used to calculate the runup amplitude, $R_{max}$.

$$\frac{R_{max}}{A_0} = \frac{2\sqrt{\pi}}{\Gamma(1+1/m)} \left(\frac{m}{m+1}\right)^{\frac{1}{2m}+\frac{1}{4}} \left(\frac{\omega^2 h_0}{g\alpha^2}\right)^{\frac{1}{2m}+\frac{1}{4}}. \qquad (29)$$

For the plane beach ($m \to \infty$) formula (29) reduces to

$$\frac{R_{max}}{A_0} = 2\sqrt{\pi} \left(\frac{\omega^2 h_0}{g\alpha^2}\right)^{1/4}, \qquad (30)$$

derived in the pioneer paper by Carrrier – Greenspan (1958).

Amplification ratio, $R_{max}/A_0$ as function of dimensionless frequency, $\omega(h_0/g\alpha^2)^{1/2}$ is displayed in Fig. 2. Taking into account the asymptotic character of the far-field formulas, we cut all values with an amplification factor less than 2; more rigorous analysis has shown that the limited value for an amplification factor is 2. As we can see, the amplification factor growth with increase of the wave frequency; as it is expected, shortest waves amplify significantly to com-compare with longest waves. The bay geometry influences on the amplification ratio also. If the cross-section varies with depth significantly, for instance when *m* < 2 (see equation (4)), amplification factors growth with the frequency quickly then the linear function. For the channel of rectangular-like shape (*m* > 40), the difference between (29) and (30) is weak. For large values of the amplification factor, the developed theory breaks due to wave breaking and we will discuss it later.

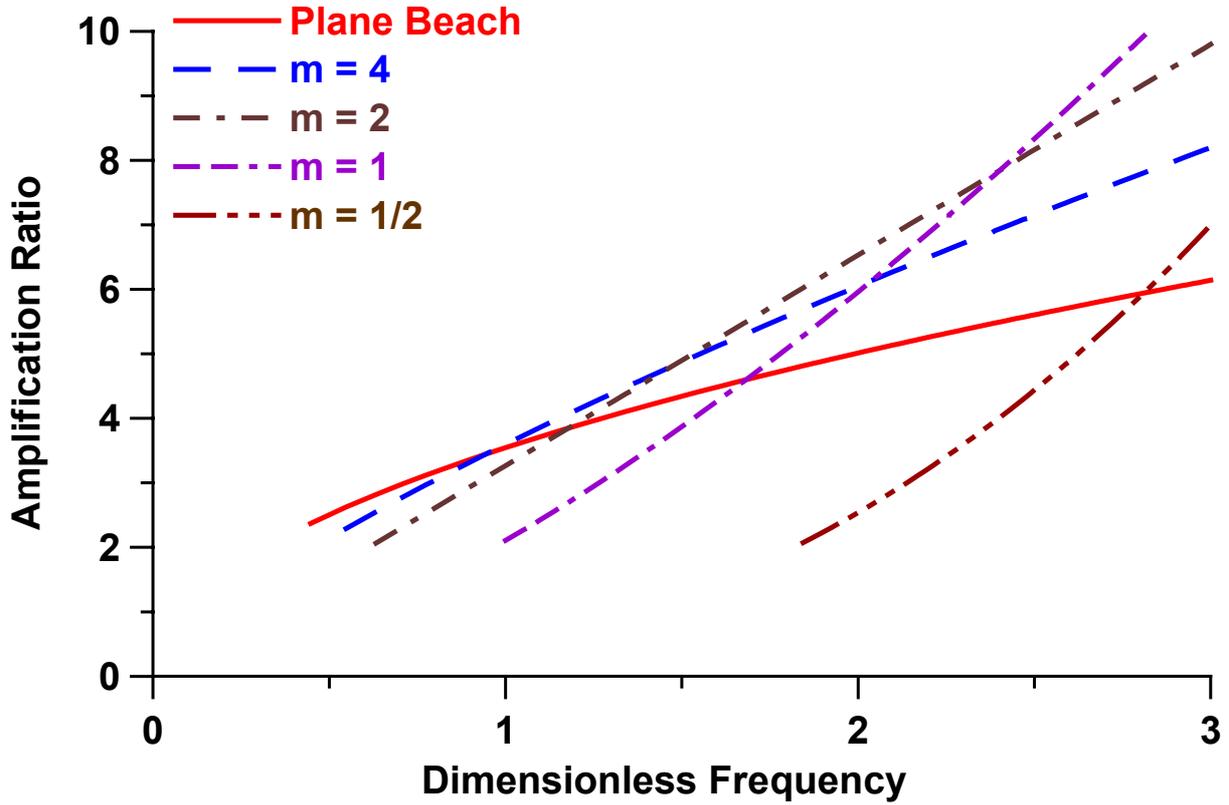

**Fig. 2.** Amplification ratio versus dimensionless frequency for sine wave climbing on a beach

From (26) the shoreline velocity in the linear approximation can be calculated explicitly

$$U_l(t_l) = -\frac{\omega R_{max}}{\alpha} \sin(\omega t). \qquad (31)$$

Now we may determine "real" nonlinear dynamics of the moving shoreline using (21) and (22)

$$U(t) = -\frac{\omega R_{max}}{\alpha} \sin(\omega(t + U/g\alpha)), \qquad (32)$$

$$\eta(t,0) = R(t) = R_{max} \cos(\omega(t + U/g\alpha)) - \frac{U^2(t)}{2g}.  \qquad (33)$$

First of all, these formulas give multi-values curves at the large wave amplitudes. For the analysis of the condition of the multiple curve appearance, the time derivative from the shoreline velocity can be calculated

$$\frac{dU}{dt} = \frac{U'_l}{1 - \frac{U'_l}{g\alpha}},  \qquad (34)$$

where prime means the derivative from "linear" velocity on its argument. Derivative $dU/dt$ tends to infinity (wave breaks) when

$$Br = 1,  \qquad (35)$$

where the parameter Br ("breaking parameter") is

$$Br = \frac{U'_l}{g\alpha} = \frac{\omega^2 R_{max}}{g\alpha^2}.  \qquad (36)$$

This condition is an another form of the zero value of the Jacobian $\partial(t,x)/\partial(I_+,I_-)$ provided by the breaking of the hodograph transformation. The form (35) is more convenient because immediately gives the maximum possible value of the runup height of tsunami waves without breaking.

Figure 3 shows the time history of the velocity and vertical displacement of the moving shoreline for various values of the breaking parameter, $Br$. Time is dimensionless ($\omega t$), and velocity and displacement are normalised on their maximum values for $Br = 1$. With increase of amplitude of the incident wave, the velocity shape tends to the shock on its front. The shape of vertical displacement is symmetrical (about vertical axis), and time of flooding exceeds the duration of the ebb phase. The shape has a jump of derivatives at $Br \to 1$, and firstly, wave breaks in the sea after deep ebb. It is important to mention that the formal description of the shoreline dynamics does not depend from the bay geometry (parameter, *m*) and determined by the runup height only; last characteristics, of course, depends strongly from the bottom geometry.

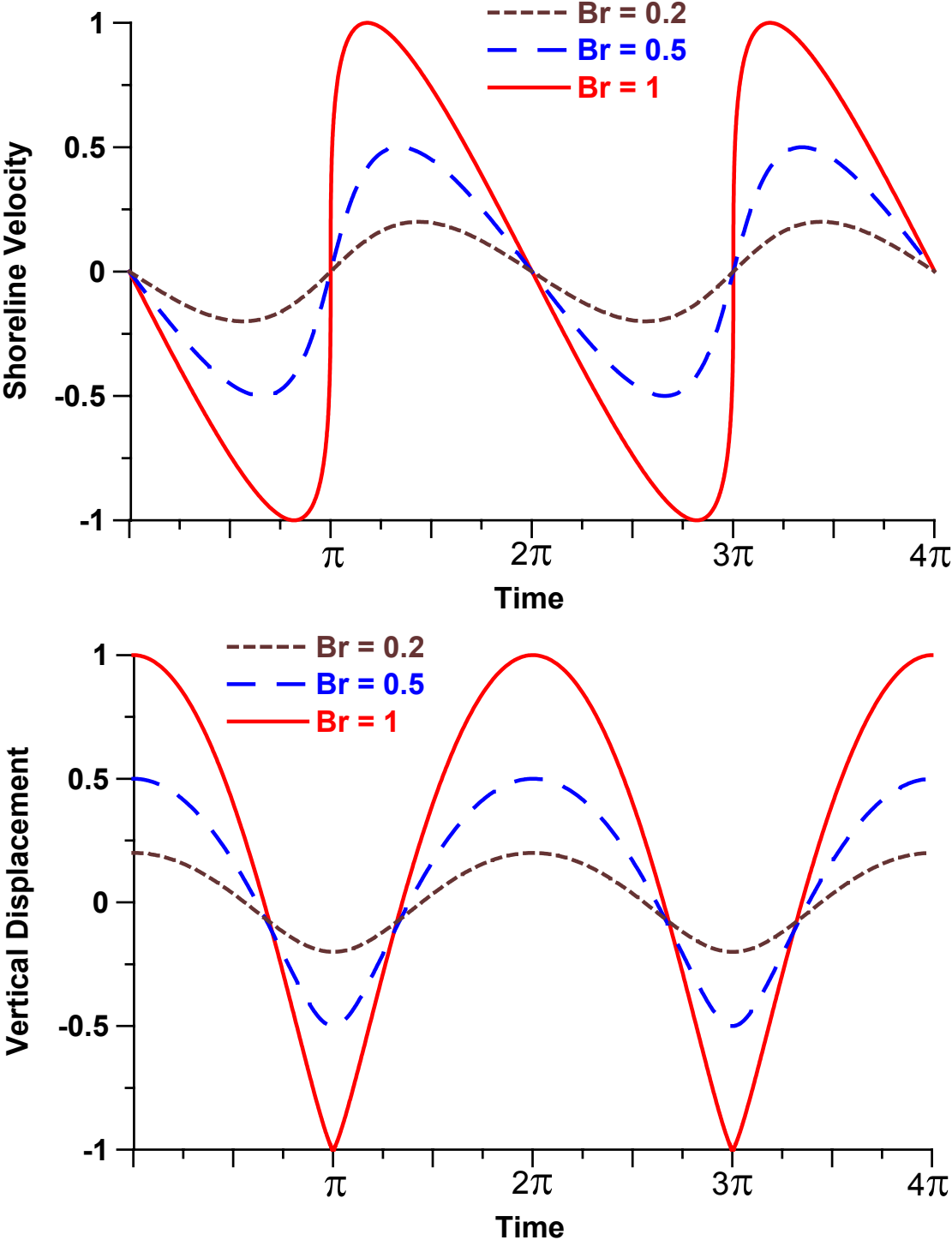

**Fig. 3.** Velocity and vertical displacement of the moving shoreline

## 5. Solitary wave runup in "parabolic" bay

Considered above sine wave can be used to study the runup of the solitary wave using the Fourier or Laplace transformation. It is important to mention one particular case of the bottom geometry when the pulse solution can be presented in close form. It is the case of the bay with parabolic cross-section ($m = 2$). The general solution of the wave equation (16) bounded everywhere is

$$\Phi(\lambda,\sigma) = \frac{F(\lambda-\sigma) - F(\lambda+\sigma)}{\sigma}, \qquad (37)$$

where $F$ is an arbitrary function. The first, the linear approximation should be used, and the vertical water displacement at any time is

$$\eta(t,h) = \left(\frac{h}{h_0}\right)^{-1/2}\left[\Psi(t - \sqrt{6h_0/g\alpha^2} + \sqrt{6h/g\alpha^2}) - \Psi(t + \sqrt{6h_0/g\alpha^2} - \sqrt{6h/g\alpha^2})\right], \quad (38)$$

where coordinate, $x$ replaced by the unperturbed depth, $h$, and the function, $\Psi(t)$ describes the time shape of the incident wave. The water oscillation on the shore ($x = 0$) is found explicitly

$$\eta(t,0) = R(t) = \sqrt{\frac{24h_0}{g\alpha^2}}\frac{d\Psi(t)}{dt}. \qquad (39)$$

If the incident wave is the solitary wave of the parabolic shape

$$\eta_{in}(t) = A_0\left[1 - \left(\frac{2t}{T}\right)^2\right] \qquad (40)$$

with amplitude $A_0$ and duration $T$ at $|t| < T/2$, and zero at $|t| > T/2$, the water on shore will vary according to

$$R(t) = -A_0\sqrt{\frac{24h_0}{g\alpha^2}}\frac{8t}{T^2} \qquad \text{at } |t| < T/2 \qquad (41)$$

with maximum amplification

$$\frac{R_{max}}{A_0} = 8\sqrt{\frac{6h_0}{g\alpha^2 T^2}} \ . \tag{42}$$

Both curves normalized on their maximum are presented in Fig. 4 with dimensionless time, *2t/T*. It is important to mention that the incident wave and the shoreline oscillation are shifted on tsunami travel time from depth $h_0$ to shore

$$t_t = \sqrt{\frac{6h_0}{g\alpha^2}} \ , \tag{43}$$

which is not shown in Fig. 4. This exact solution demonstrates that tsunami wave can appear very quickly, then water will recede to the sea during time of *T*, and then again quickly returned to the initial state. Such behaviour of tsunami waves very often pointed in observations. The short time of the tsunami approach is related with finiteness of the initial disturbance, and can explain the appearance of two characteristic times in tsunami waves, when the initial disturbance has only one time scale. In fact, our solution is not smooth, and it should immediately breaks. Smoothing the wave shape in the "end" points, we may obtain the smooth oscillation of the shoreline but it will continue to keep two characteristic times. The appearance of the second scale is related with frequency properties of the shelf amplifying the high spectral harmonics more significantly. Because finite disturbances are used very often numerically, it is necessary to check physical reliability of the wave behaviour in the 'end" points.

It is necessary to mention that previously in tsunami literature, the shape of the initial disturbance in the tsunami source was chosen in the form of elevation (crest) of the water displacement. Now, the dipole-like shape of the initial wave is preferable (Tadepalli and Synolakis, 1996; Synolakis et al, 1997). For the parabolic shape of bay (*m = 2*), as it is shown above, the beach "differentiates" the incident wave, and the runup of each individual wave can be considered independently, and here there is no collective effect of individual wave interaction. Approaching of the crest only induces the tide and then the ebb (if the incident wave is depression, it induces the receding and then flooding). Thus, the problem of the runup of the dipole-like wave is solved trivially. For other geometries in general case, the runup of each individual wave (even it is finite) accompanies by the long-time tail, and the runup dynamics can be more complicated. This process investigated in details now for the plane beach only (Tadepalli and Synolakis, 1994).

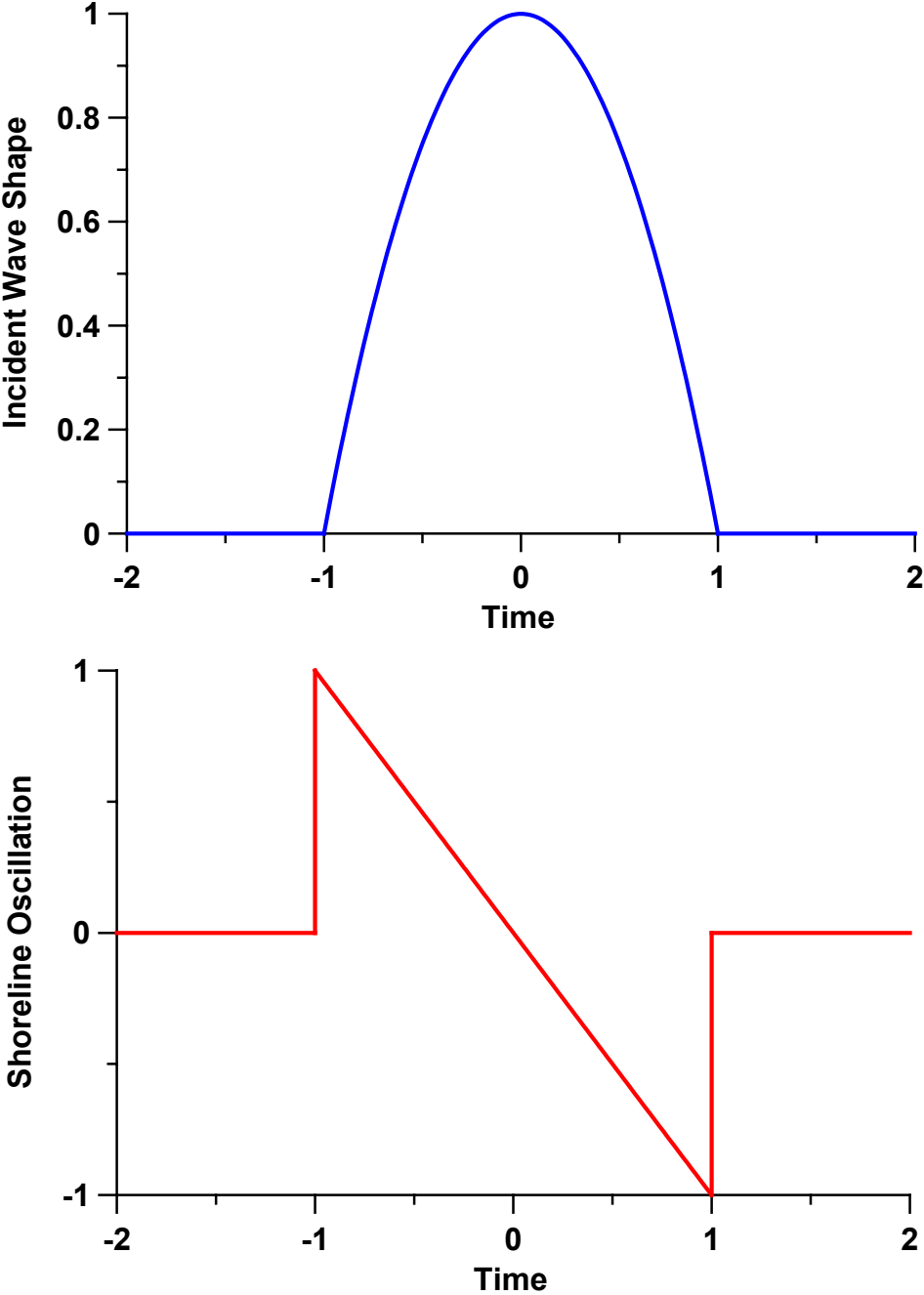

**Fig. 4.** Wave transformation in a bay of parabolic cross-section

# 6. Conclusion

To compare with previous studies the given paper analyses the runup of tsunami waves on the coasts of the barrow bays, channels and straits. Using the narrowness of the water channel, the one-dimensional equations are applied of the nonlinear shallow-water model; they include the variable cross-section of channel. The specific shape of the coastal zone with parabolic-like cross-section is considered to achieve the analytical solutions with use of the hodograph (Legendre) transformation similar to the Carrier and Greenspan transformation for the wave runup on the plane beach. As a result, the linear wave equation is derived and all physical variables (water displacement, fluid velocity, coordinate and time) can be determined through the solutions of this linear equation. The dynamics of the moving shoreline (boundary of the flooding zone) is investigated in details. It is shown that all analytical formulas for the moving shoreline can be obtained explicitly in two steps for tsunami originated far from the coast. On the first step, the linear approximation is used to calculate the "linear" water oscillations on the shore. On the second step, these linear expressions converted in "nonlinear" formulas for the moving shoreline with nonlinear transformation of the time axis. The condition of the wave breaking is derived; it includes maximum wave amplitude on the shore, wave frequency and face-beach slope. Two examples of the incident wave shapes are analysed: sine wave and parabolic pulse. The last example demonstrates that weak singularities on the end of initial disturbance can lead to strong singularities in the dynamics of the moving shoreline due to frequency properties of the coastal zone. Even in the case when the one crest only approaches to the shore, the flooding can appear very quickly; then water will recede relatively slowly, and then again quickly return to the initial state.

The numerical codes to simulate the tsunami wave runup are developed and applied for specific case studies, see for instance (Titov and Synolakis, 1997; Choi et al, 2003), Unfortunately, the computations of the tsunami wave run-up and inundation require large computer resources. To estimate the tsunami risk requires to perform more than 100 tsunami scenario (Zahibo et al, 2003) and it is really difficult to compute wave runup in each variant. It was shown that the analytical formula for plane beach (30) leads to best comparison with observations than the "vertical wall" approximation usually applied in such calculations (Choi et al, 2002). We hope that analytical "runup" formulas given above for basins of more complicated geometry can be useful for correction of the results of "non runup" calculations.

*Acknowledgements*

This study is supported particularly by grants from EGIDE, INTAS (01-2156 and 03-51-4286) and RFBR (05-05-64265).